# Towards the integration of formal specification in the Áncora methodology

Carlos Alberto Fernández-y-Fernández [1] y Martín José José [2]

[1]Instituto de Computación, Universidad Tecnológica de la Mixteca, carretera a Acatlima Km. 2.5 Huajuapan de León, Oax., México C.P. 69000,
caff@mixteco.utm.mx

[2]División de Estudios de Posgrado, Universidad Tecnológica de la Mixteca, carretera a Acatlima Km. 2.5 Huajuapan de León, Oax., México C.P. 69000,
martinjj2009@gmail.com

**Abstract:** There are some non-formal methodologies such as RUP, OpenUP, agile methodologies such as SCRUP, XP and techniques like those proposed by UML, which allow the development of software. The software industry has struggled to generate quality software, as importance has not been given to the engineering requirements, resulting in a poor specification of requirements and software of poor quality. In order to generate a contribution to the specification of requirements, this article describes a methodological proposal, implementing formal methods to the results of the process of requirements analysis of the methodology Áncora.

## 1. Introduction

Generating quality software has been a problem for many years for the software industry. In the late 60's the computer community detected a number of cases of failure, such as late delivery of the final product exceed the original cost, poor product quality. In late 1968, government, academia and users agreed to the first conference on Software Engineering (SE) to identify methods and techniques for software development quality and best cost [1]. Some authors such as Sommerville [2] define SE as "*a discipline of engineering that incorporates all aspects of software production from the early stages of system specification up to its maintainance and after it is used*".

Over the years some methodologies such as RUP [3], OpenUP [4] and agile methodologies like SCRUP [5], XP [6] for the software development process have been defined. However, T.E. Bell and T.A. Thayer [7] Lamsweerde [8] proposed strengthening SE Requirements Engineering (RE), to review the requirements phase, because it is considered the basis for any software. Lamsweerde [8] defines the RE as "*a coordinated set of activities to explore, evaluate, document, consolidate, revise and adapt the goals, capabilities, qualities, limitations and assumptions that the system can meet on the basis of the problems and opportunities offered by new technologies.*" One way to structure the process of RE is elicitation, analysis, specification, validation and management of software requirements [9] with the aim of obtaining clear, unambiguous, consistent and compact requirements.

The most relevant phase for the research about to be carried out is the Software Requirements Specification (SRS), in which it is set in writing what the new system is expected to do and not to do. It is also considered the basis for an agreement between the customers and suppliers (developers) [10]. One way to eliminate inconsistency and ambiguity in the results of the specification document is through formal methods, which involves using formal specification languages [11]. The purpose of this proposal is the use of formal methods in SRS.

## 2. The research environment

At the beginning of a software development, the customer needs, should be clear, the benefits of the new product should be considered, if the software solves the problems faced by the customer, and this involves understanding the application area, the operational constraints of the system, the specific functionality required by the client and the essential characteristics of the system, such as performance, safety and reliability [12]. Then to generate a quality software it is important to use techniques, methods and methodologies. The article describes a formal methodological process, but it is important to look at some concepts which have been considered the fundamental principles of the study area. The following paragraphs describe briefly each one of them.

The proposal focuses on the SRS of the RE. An SRS is the documentation of all the requirements selected in the phases of elicitation and analysis. Bruade [13] conceptualizes the SRS as functional specific requirements and belonging to the developer and defines them as *"a list of specific properties and functions which the software must have in order to provide the desired service, the restrictions of operation and its development, explained in detail based on the needs obtained in the elicitation phase and the selection made in the analysis."*

Montoya [14] defines formal methods (MF) as a series of techniques of mathematical logic to specify, design, implement and verify the information systems through a specification language. This specification language is structured by a vocabulary, syntax and semantics based on set theory, logic and algebra of predicates [8].

In the past has been said that the use of a methodology is important for software development. In this case, Áncora, a methodology for analyzing software requirements conducive to reuse will be used [15] [16]. Áncora facilitates the generation of requirements analysis, its techniques and tools combine methods from several branches of knowledge such as planning, social psychology, computational linguistics and SE. The Fig. 1 shows an outline of general activities that take place with Áncora and some artifacts that this produced. The central element is a set of scripts and dialogues that represent the requirements model and through them occurs view integration, the translating for some desired development patterns (eg., object-oriented, data flow, states) that feeds and is fed by reuse elements of similar systems.

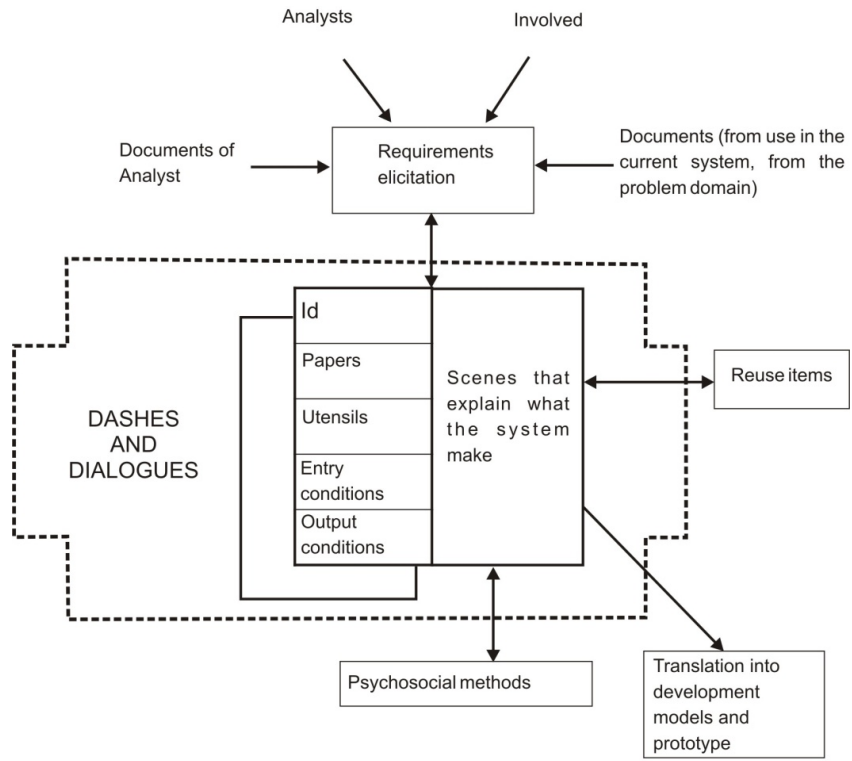

**Fig. 1** Scheme of elements of requirements analysis with Áncora [15]

The last element to describe is the Eclipse Process Framework Composer (EPF Composer) [17]. The EPF Composer is a free tool, its development is based on the Eclipse platform, it uses as languages of modeling software processes the structure of the Software & Systems Process Engineering Meta-Model Specification 2.0 (SPEM 2.0) and the Unified Method Architecture (UMA). Both have provided benefits for the description of development methods and processes. Besides that, EPF Composer supports different methods for the lifecycle of a project, such as Cascade models, incremental and iterative development. Due to these characteristics, EFP has become one of the most representative models of processes of software [18].

## 3. Defining the problem

The company Group Standish [19] [20] has generated descriptive reports called Chaos Report, where it shows that from 2000 to 2009 most of the projects do not meet user requirements, which remains a big problem in the software industry, meaning that most software development companies do not have adequate knowledge about the requirements or do not use a methodology for their developments.

The reports provided by Group Standish, indicate that many of the problems of software engineering comes from a poor specification of requirements. For example for a system developer it is common to make ambiguous interpretations of a requirement in order to simplify its implementation. However, there are actions that hamper the development because they fall outside the context of the customer's needs. Such decisions allow the stipulation of new requirements resulting in changes to the system, delaying the delivery of the product and increaseing the cost stipulated. In recent years there have been various types of failures in software development [1] [21], such as:

- The low quality of the projects due to the lack of quality metrics and the use of quality control tools,
- The time and the budget is exceeded in the development of software projects due to poor planning, and
- The high cost in personnel for development and maintenance.

## 4. Related works

In order to gather information that allows an approach to the subject of study and to learn about other similar cases, detailed below are some research studies which have been done and have something in common with the proposal, besides the fact that they will be useful for the activities which are to be carried out.

### 4.1. Formal specification in OCL of consistency rules between class diagrams and UML use cases and model interfaces.

Research carried out by Zapata and Gonzales [22] mentions the importance of an specification process of requirements, in which customer's needs are interpreted and a solution is offered, commonly using the UML tool. The paper describes that currently in the software industry this process is carried out in an informal way and possibly vaguely, which is why it is common for inconsistency to exist between a pair of diagrams.

The proposal of the authors is to establish a method to verify the relationship between the class diagram [23], the use case diagram of Unified Modeling Language (UML) [23] and the graphical user interfaces [24]. The authors state 8 rules of consistency and assure accuracy of information provided by the language models using the Object Constraint Language (OCL). As a result of his research is to obtain the definition of a formal specification of consistency rules, thus allowing that analysts do not invest too much time and money in the comprehensive review of the diagrams and detecting errors in the early stages development.

### 4.2. Specification of contracts of software using OCL

In the research that establishes Vignaga [25], it allows to encourage the use of formal methods and shows that it is possible to use OCL as a language of specification formal of contracts of software. The aim of his research is to show the importance and usefulness of OCL for specifying pre-and post condition in the software contracts, focused on object-oriented structural modeling and use case diagrams use. The author defines that the essence of a software contract is knowing the state that an instance must have (on which the operation is applied) before the execution of the operation (preconditions), and describing the state of the instance after the execution of the operation (post conditions). This has control of operations and states, based on predefined rules.

González and Zapata used the formal language to verify the consistency of non-formal techniques (class diagrams, use cases); Vignaga implements the formal language as part of a requirements specification document. Both works offer an incomplete methodology for developing a formal specification of requirements. The proposal of our work is to provide a complete and documented methodology, as described in the following topics.

## 5. Work proposal

The proposal complements the life cycle of Áncora incorporating a formal specification in the stage of "*conflict resolution, prioritization and validation of requirements*"; making a contribution to the IR and knowing the benefits of an FM. Based on the foregoing, during the development of the research we seek to verify whether it is possible to implement a formal requirements specification in the stage mentioned.

The issue has been raised because it has been investigated that a formal specification through logic and mathematical techniques is an effective means to assess the requirements, to assure quality in the systems in non-trivial complexity, to avoid costly corrections after the development [26]. The methodological procedure to this research is comprised as follows:

- To generate a conceptualization of the history of the topic, through a systematic review of the literature.
- To collect background information on the contributions of formal methods in software development.
- To analyze the requirements specification phase in different methodologies for software development.
- To describe methods and techniques for software requirements specification.
- To investigate and analyze some formal specification languages in order to choose one or more to be used in the proposal.
- To define the formal specification of methodology Áncora using formal specification languages, some rules and standards referred to in the RE.
- To generate an evaluation of the proposal using available tools that support formal languages used.
- To document the processes in EPF Composer.

## 6. Approach to the solution

The overall research objective is "to integrate the formal aspects for the requirements specification phase of software in the Áncora methodology," considering the activities set to generate the analysis of software requirements. The proposal consists in using the specification phase of the RE and using one or more formal languages, rules and standards, proposing activities to build a methodology that allows to define a formal specification of requirements. So far we have considered using tasks algebra [27] [28] alloy [29] [30] and OCL [31] as formal specification languages. The proposed development has been defined in four stages, represented by a dotted line rectangle (*see Fig. 2*).

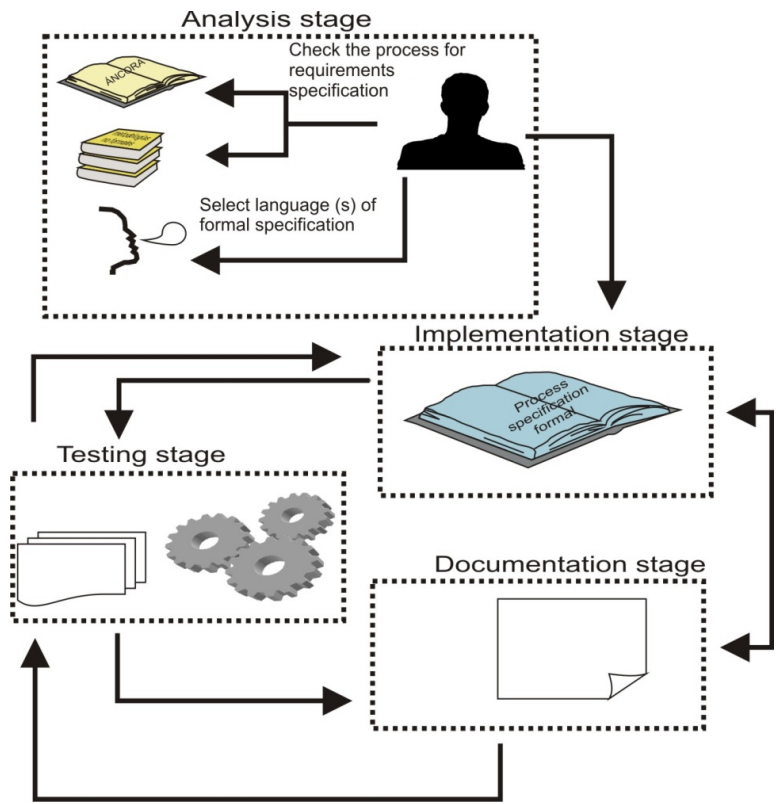

**Fig. 2** Stages of the proposal

## 7. Conclusions

Technological advances, new complex needs that the human being has and the economic competition have put pressure on the software development process. However, the software industry has a serious problem to generate a quality product.

This is due to lack of interest in applying the process of RE, causing projects to fail because some requirements are not fulfilled or are incomplete. The research is responsible for strengthening the requirements specification, defining activities to generate a consistent and unambiguous specification, through a proposal that is effective due to the use of formal methods and formal specification languages whose vocabulary, syntax and semantics are formally defined and based on mathematical concepts from set theory, logic and algebra of predicates. This form is expected to generate a complement to the methodology Áncora and a documentation standard for software development.